\documentclass[12pt]{article}

\usepackage{scicite}
\usepackage{graphics,epsfig}

\usepackage{times}

\newenvironment{sciabstract}{%
\begin{quote} \bf}
{\end{quote}}

\newcounter{lastnote}

\title{Failure mechanisms of graphene under tension}

\author
{Chris A. Marianetti$^{1\ast}$ and Hannah G. Yevick$^{2}$  \\
\\
\normalsize{$^{1}$Department of Applied Physics and Applied Mathematics, Columbia University}\\
\normalsize{$^{2}$Department of Physics, Columbia University}\\
\normalsize{500 West 120th St., 1144 Mudd Hall, New York, NY 10027, USA}\\
\\
\normalsize{$^\ast$E-mail:  chris.marianetti@columbia.edu.}
}

\date{}

\begin{document} 


\baselineskip24pt


\maketitle


\begin{sciabstract}
Recent experiments established pure graphene as the strongest material known to
mankind, further invigorating the question of how graphene fails.  Using
density functional theory, we reveal the mechanisms of mechanical failure of
pure graphene under a generic state of tension.  One failure mechanism is a
novel soft-mode phonon instability of the $K_1$-mode, whereby the graphene
sheet undergoes a phase transition and is driven towards isolated benzene rings
resulting in a reduction of strength.  The other is the usual elastic
instability corresponding to a maximum in the stress-strain curve. Our results
indicate that finite wave vector soft modes can be the key factor in limiting
the strength of monolayer materials.

\end{sciabstract}

The mechanical failure of materials is usually a complex process which may
involve defects at a variety of length scales, such as dislocations, grain
boundaries, cracks, etc. The complexity and statistical nature of these defects
cause mechanical failure to be extremely dependent on not only the type of
material, but also on the manner in which the material was synthesized.  To the
contrary, ideal strength, which can be defined as the maximum attainable stress under a
uniform strain field in the absence of any instabilities, is an intrinsic property of a
material\cite{Zhu2009167}. 
Perhaps the simplest instability is the so-called elastic instability, whereby a maximum
in the stress-strain relation is achieved while retaining the symmetry of strained lattice. 
There has been a significant experimental effort throughout the past
century to prepare sufficiently pure samples such that the ideal strength could
be probed, including early experiments with fine glass rods and steel
wires\cite{Orowan1949185}.  Recently, the measurement of ideal strength has
been achieved in the case of graphene\cite{Geim2007183}, a monolayer of carbon. Using nanoindentation, Lee et al
managed to strain graphene until failure under conditions which appear to be
very nearly ideal\cite{Lee2008385}. This experiment reinvigorates the
fundamental question of how and why a material fails under ideal conditions.
The answer lies within the forces which bond a material together. Computing
these forces from the first-principles of quantum mechanics is made possible by
intelligent approximations to the quantum many-body problem, such as the local
density approximation (LDA) of density functional theory
(DFT)\cite{Hohenberg1964B864,Kohn19651133}, in addition to plentiful
computational resources.  While LDA may qualitatively break down in certain situations where the
electronic correlations are strong\cite{Kotliar2006865}, it works extremely
well in materials with relatively large electronic bands such as graphene.
Therefore, one can reliably explore the mechanical properties of graphene
from first-principles. In this study, we use DFT to determine the mechanism of
mechanical failure for an arbitrary state of tension.

To determine the elastic instability of a material,
DFT can be used to generate the forces as a
function of strain, and
such studies were performed once sufficient computational power was available (see reference \cite{Morris20002827} and 
references therein). However,
there is no guarantee that the structure will remain stable with respect
to \emph{inhomogeneous} deformations
under strain. Indeed, a structure may become unstable and transform to a
different structure with a lower ideal strength.  In order to determine if a
structure is mechanically stable, one needs to consider lattice vibrations and  confirm that all of the phonon
energies are real and positive\cite{Dove1993book}. 
If any of the phonon energies were to
go to zero, there would be no energy barrier for the crystal to distort along
the zero-energy phonon mode.  Therefore, the phonon spectrum is the fundamental
entity which dictates structural stability. Phonon modes with zero or very
small energies, excluding the acoustic phonons for $k\rightarrow 0$, are usually termed "soft-modes"\cite{Axe197332}. There are
numerous structural phase transitions in which the two phases are directly
connected by a soft-mode, and the concept of the soft-mode gained prominence in
the context of elucidating the ferroelectric transition in
BaTiO$_3$\cite{Cochran1960387,Ginzburg20011037}.  In this work, we 
demonstrate that a soft-mode is responsible for a phase transition and the
resulting  mechanical failure of graphene in certain states of tension.
Previous DFT studies of \emph{bulk} systems such as Ni\cite{Clatterbuck2003135501} and Si\cite{Dubois2006235203} have 
demonstrated that non-trivial acoustic phonon instabilities may precede the usual 
elastic instability for certain states of strain and therefore limit the ideal strength of the material.
However, these scenarios are extremely difficult to decipher experimentally, even
indirectly, due to defects and plastic deformation, while our predictions in
graphene may be directly tested experimentally.

In the case of graphene, previous phonon calculations have determined that the
elastic instability is the mechanism of failure for uniaxial strain in the
armchair or zig-zag directions\cite{Liu2007064120} (ie. $x$ and $y$ directions
in panel C/D/E of Figure 2, respectively). However, the mode of failure in a more
general state of strain has never been considered. In this work, we computed
the phonons using the displacement method\cite{Walle200211}, where the forces
are generated using DFT.  All DFT calculations were performed using the Vienna
ab-initio simulation program
(VASP)\cite{Kresse1993558,Kresse199414251,Kresse199615,Kresse199611169}, which
uses a plane-wave basis and projector-augmented wave (PAW)
potentials\cite{Kresse19991758}.  An energy cutoff of $375 eV$ was used with
the soft VASP carbon PAW, and k-point meshes of $27\times27$, with respect to
the primitive cell. A state of strain was constructed by applying
the strain and allowing all cell internal coordinates to relax. Supercells of
$9\times9$ times the primitive cell were used when computing the force
constants. The displacement method involves computing the real space force
constants between atoms by displacing an atom and evaluating the resulting
forces.  Previous work has established that the displacement method is accurate
for unstrained graphene\cite{Dubay2004089906}. In Figure 1 we present the
phonons for equibiaxial strains ranging from zero to
$\epsilon_A=(\epsilon_x+\epsilon_y)/\sqrt{2}=0.212$, where $\epsilon_i$ is the
nominal strain $\ell_i/\ell_{io}-1$.  In Figure 1A, we reproduce the phonons
for unstrained graphene, showing excellent agreement with previous
work\cite{Dubay2004089906,Liu2007064120}.  
Due to symmetry, all phonon modes contain purely either out-of-plane or in-plane
atomic displacements.  Three acoustic branches  and three optical branches are
present, and the $z$-axis acoustic branch has a nonlinear dispersion at small
wave vector as expected for a monolayer.  In Figure 1B, the phonons are shown for
the case of an equibiaxial strain of $\epsilon_A=0.106$.  A significant
softening of the in-plane phonons is observed, which is to be expected given
that all the in-plane distances are increasing uniformly.  The out-of-plane
phonons change less in general, and one notable change is that the quadratic
dispersion of the acoustic out-of-plane mode becomes closer to linearity.   In
Figure 1C, corresponding to a strain of 
$\epsilon_A=0.205$, it becomes clear that certain phonon modes are softening at
a much higher rate than others. In particular, the $K_1$ mode at the  $K$-point
 is rapidly dropping towards zero. Upon reaching a strain of
$\epsilon_A=0.212$ (see Figure 1D), the $K_1$-mode has become imaginary resulting in a
soft-mode.  This implies that the structure has become unstable and will
undergo a phase transition by distorting along the $K_1$-mode.  Group theory
alone dictates the nature of this $K_1$-mode, and by considering 
linear combinations of both $K$ and $K'$
one arrives at two distinct real distortions\cite{Basko2008041409} (see figure 2C/D/E).
These modes can be classified as the $A_1$ and $B_1$ irreducible representations of the $C_{6v}$ point group. 
The positive amplitude of the $A_1$ mode can be viewed as the formation of individual
benzene rings where all edge-sharing benzene neighbors have been annihilated, while both the $B_1$ and
negative $A_1$ amplitudes form dimerized structures.
While the $A_1$ and $B_1$ modes transform differently under $C_{6v}$, these modes form a 2-fold representation when
including the lattice translations. 
Below we show the positive $A_1$ mode is most energetically favorable when including anharmonicity.

Given that the $K_1$-mode has a reasonably short wavelength, one can directly
explore the properties of this mode with a 6-atom unit cell, which is three
times the size of the primitive cell (see Figure 2C/D/E).  We shall refer to
this enlarged unit cell as the $K$-cell hereafter. The energy is computed as a
function of the amplitude of the $A_1$-mode and $B_1$-mode at a series of different equibiaxial strains (see
Figure 2A/B). 
The curvature at zero amplitude (ie. $k/2$) 
gives the phonon energy $\hbar\sqrt{k/m}$, which must be identical for these degenerate modes.  
As the strain is increased, the modes continually becomes softer and eventually
the curvature at zero amplitude goes to zero and the modes becomes soft
simultaneously.  This analysis predicts the soft mode to occur at $\epsilon_A=0.213$,
independently confirming the results of our phonons which yield a soft mode at
$\epsilon_A=0.205-0.212$.  Further strain results in a double-well potential, where the well depth
and amplitude increases with increasing strain.  In the absence of
anharmonicity, one would find the well known "Mexican hat" potential 
characteristic of a two-fold representation, but anharmonicity results in the
more usual "warped Mexican hat" potential (see Figure 2A inset). As shown in Figure 2, the
minimum energy corresponds to the positive amplitude of the pure positive $A_1$ mode
(or identically $-\frac{1}{2}A_1\pm \frac{\sqrt{3}}{2}B_1$  ).  This is physically reasonable given that the
positive $A_1$ mode forms benzene rings while the other distortions result in
dimerized structures.

It should be emphasized that at this point one does not know if the material
will fracture, only that a phase transition will occur. Therefore, we must
explore the strength and stability of this new phase of strained graphene. 
 The stress as a function of the
equibiaxial strain is computed for both the primitive unit cell and the
$K$-cell (see Figure 3). For the primitive cell, the curve is smooth and the
elastic instability occurs at a strain of 
$\epsilon_A=(\epsilon_x+\epsilon_y)/\sqrt{2}=0.307$.  However, the primitive unit
cell does not have the freedom to distort along the $K_1$-mode as the primitive
translational symmetry is enforced in the calculation (ie. the primitive cell is smaller
than the wavelength of the $K_1$-mode). The same curve can now
be analyzed for the $K$-cell. The phonon-instability is clearly illustrated
by a discontinuity in the curve at $\epsilon_A=0.213$ (see inset of Figure 3), in excellent agreement
with our preceding two calculations. Upon activation of the
$K_1$-mode, the force rapidly drops, and is subsequently nearly flat
until decreasing.  Therefore, this new phase which forms is essentially
mechanically unstable, and there is no need to recompute the phonons for this
new phase. As a result, the soft $K_1$-mode can be seen not only as the precursor
to a phase transition as in soft-mode theory, but also as a soft-mode which
leads directly to mechanical failure. 

The above analysis has revealed that for equibiaxial strain the mode of failure
of graphene is radically different than the usual elastic instability which is
observed for uniaxial strain in the zig-zag or armchair directions. Therefore,
the question arises as to when the elastic instability is the failure mode
versus the $K_1$-mode instability for a generic state of tension.  In order
to resolve this we have computed the stress versus strain for both the primitive unit cell and the $K$-cell, as above, 
for all possible linear combinations of tensile
strain in the zig-zag and armchair directions (see Figure 4A). This
two-dimensional space of strains can be characterized in polar coordinates with
an angle and a radius, and $\theta=45$ corresponds to equibiaxial strain.  As shown in Figure 4, the plot is naturally separated
into three regions. In the first and third region the elastic instability
precedes the $K_1$-mode instability, and therefore the failure mechanism is
the elastic instability.  On the contrary, in the second region the $K_1$-mode instability
occurs first and therefore limits the strength of the material. Therefore,
strains near uniaxiality fail via the elastic instability while strains near
equibiaxiality fail via the $K_1$-mode instability. It is worth noting the degree of 
symmetry in this plot. The $K_1$-mode instability is nearly invariant for a mirror line placed
at $\theta=45$, despite the fact that there is no such lattice symmetry. Alternatively, the elastic curve has some observable
differences. The stress in the $x$ and $y$ directions at failure (ie. along the $K$-cell line in Figure 4A) is plotted as a function 
of $\theta$ (see Figure 4B), indicating the stress necessary to realize the state of strain at failure.

This analysis is not yet exhaustive due to the fact that graphene is
anisotropic, and therefore shear strain would have to be included in the
present coordinate system to enumerate every possible state of tension.
Alternatively, one could repeat the above analysis for every possible rotation
of the coordinate system which is not generated by a member of the point group
of graphene. This corresponds to generating Figure 4A for
every possible rotation of the coordinate system between zero and 15 degrees.
We have rotated the coordinate system in three degree increments and
regenerated Figure 4 at each increment (see Figure 4C for $15$ degree rotation).
Clearly, $\theta=45$, which corresponds to equibiaxial strain, will be
invariant to any rotation. Conveniently, all of the results for the different
rotations are bounded by the envelope curves created by superimposing Figure 4A
and 4C.   All rotation curves
progress monotonically with rotation between the limits of the envelope. 
The envelope for the $K_1$-mode instability is extremely narrow,
reflecting the insensitivity to rotation (ie. shear). It should be noted that
Figure 4C is symmetric about $\theta=45$ due to a mirror line which maps $x^{\prime}\leftrightarrow y^{\prime}$.

Our prediction of the soft $K_1$-mode may be directly verified
experimentally by measuring the phonon dispersion as a function of strain.
Neutrons have commonly been used to measure phonon dispersion as a function of
temperature, allowing the detection of soft modes as precursors to phase
transitions in a variety of different \emph{bulk} systems\cite{Shirane1974437}.
In the case of a \emph{monolayer}, measuring the phonon spectrum is a much more
difficult venture as not enough material is present to use neutrons.  Electrons
have been used to measure the surface phonons of graphite, which is an
excellent approximation to graphene, both using reflection electron-energy-loss
spectroscopy\cite{Oshima19881601} and high-resolution electron-energy-loss
spectroscopy\cite{Siebentritt1998427}. Therefore, the phonons could potentially
be measured directly for graphene. The challenge in this particular case would
be the fact that the graphene would have to be strained in-situ. 
Another more indirect probe would be Raman spectroscopy\cite{Basko2008041409}, which 
has already been performed for graphene under uniaxial tension\cite{Huang20097304}.

It is instructive to make a comparison of our results with
existing experimental observations. The experiments of Lee et al suspended
graphene over cylindrical holes in a SiO$_2$ substrate and used an atomic force
microscope to impinge upon the graphene until failure occurred\cite{Lee2008385}.
They assumed that the strain in the sample beneath the indenter tip
may be approximated as equibiaxial. The stress was fit to second order in the
equibiaxial strain and resulted in a Lagrangian breaking strain of
$\epsilon_x=\epsilon_y=0.250$, which corresponds to a nominal strain
of $\epsilon_x=\epsilon_y=0.225$. Unexpectedly, this far exceeds the breaking
strain as dictated by the $K_1$-mode of $\epsilon_x=\epsilon_y=0.151$.
Therefore, it is clear that theory and experiment are not operating under
identical conditions, and it is necessary to detail all significant differences.
Firstly, our calculations are performed at zero-temperature, while the
experiments are performed at room temperature. Secondly, the experiment could
be influenced by the presence of the nano-indenter tip or other elements which
may react with the graphene layer. Finally, the experiment is
assumed to be in a state of equibiaxial strain while our calculations are by construction.
Any and all of these differences may be linked to the 
difference between theory and experiment.  Interestingly,
the results of Lee et al are in reasonable agreement with our results for the
elastic instability of equibiaxial strain (ie. $\epsilon_x=\epsilon_y=0.216$).
This is suggestive that perhaps somehow the $K_1$-mode is being stabilized
in the nanoindentation experiment. One possibility is that the $K_1$-mode is
being stabilized by temperature, as the experiments are performed at room
temperature. Given the well depth of $175meV$
of the $K_1$-mode for a strain of
$\epsilon_x=\epsilon_y=0.22$  (see Figure 2), it appears
unlikely that temperature alone will account for this difference. However,
temperature will clearly influence when the $K_1$-mode activates. This issue
can be resolved by bridging theory and experiment in future work.
Nanoindentation experiments may be performed at low temperatures, and molecular
dynamics simulations may be performed at high temperature and in a geometry
similar to experiment. 

In conclusion, we have determined the failure mechanisms of graphene under
ideal conditions using DFT calculations. Graphene fails via the usual elastic
instability for any uniaxial strain. In the case of equibiaxial strain, graphene
exhibits a novel soft-mode phonon instability at the $K$-point which results in
a phase transition into an unstable phase and leads to mechanical failure.
This $K_1$-mode instability corresponds to the graphene sheet distorting towards
isolated benzene rings, and it goes soft for
$\epsilon_A=(\epsilon_x+\epsilon_y)/\sqrt{2}=0.213$ which significantly
precedes the elastic instability at $\epsilon_A=0.307$. We have mapped out the
elastic failure and $K_1$-mode failure for all possible tensile strain
states, documenting the transition between the two modes. Further experiments
have been suggested to directly test our prediction of the softening of the
$K_1$-mode.

We gratefully acknowledge support from NSF via Grant No. CMMI-0927891. We thank J.W.
Kysar, M.E. Manley, and J. Hone for fruitful conversations.

\bibliographystyle{Science}
\bibliography{science}

\clearpage

\noindent {\bf Fig. 1.}({\bf A-E}) The phonons of graphene for different levels of equibiaxial strain. Blue lines indicate the phonons
within the plane while red lines indicate phonons out of the plane. The $k$-point
labels $\Gamma,M,K$ correspond to \\
$(0,0),(0.5,0),(1/3.,1/3.)$, respectively,
in fractions of the reciprocal lattice vectors. A primitive cell of 
$\vec{a}_1=(a\sqrt{3}/2,-a/2), \vec{a}_2=(0,a)$ was used, where $a$ is the nearest-neighbor bond length.
The different panels correspond to various levels of equibiaxial (ie. $\epsilon_x=\epsilon_y$) strain 
$\epsilon_A=(\epsilon_x+\epsilon_y)/\sqrt{2}$. A black arrow is used to
identify the $K_1$ mode.  A soft-mode can clearly be identified at the
$K$-point for  $\epsilon_A=0.212$.

\noindent {\bf Fig. 2.} ({\bf A}) The energy as a function of the $A_1$ phonon
amplitude for equibiaxial strain $\epsilon_A=(\epsilon_x+\epsilon_y)/\sqrt{2}=0-0.311$ in increments
of $\sqrt{2}/100$.    The line color changes from green to yellow to red as strain increases. The phonon goes
soft between the 15th and 16th curve corresponding to a strain of $\epsilon_A=0.212-0.226$. The inset of panel
A is a contour plot of the energy vs the $A_1$ and $B_1$ phonon amplitudes for a strain of $\epsilon_A=0.311$.
({\bf B}) The repeat of panel A for the $B_1$ phonon.
({\bf C,D,E})
The negative $A_1$, positive $A_1$, and $B_1$ phonons, respectively ($B_1$ mode is symmetric). The undistorted lattice is shown in grey. 
The unit cell of the distorted structure (ie. the $K$-cell) is denoted with dotted purple lines. For illustrative purposes, the amplitudes
shown corresponds to 2.5 times the amplitude for the respective well minima and $\epsilon_A=0.311$.

\noindent {\bf Fig. 3.} Nominal stress vs. equibiaxial strain for a primitive unit cell
(blue line) and the $K$-cell (red line). The inset shows a magnified view of
where the $K_1$-mode goes soft, as indicated by a discontinuity in the curve. 

\noindent {\bf Fig. 4.} ({\bf A}) The maximum stable strain for the primitive
unit cell (blue curve) and the $K$-cell (red curve) as a function of all
possible linear combinations of zig-zag and armchair uniaxial tensile strains.
A given direction of strain corresponds to an angle $\theta=0-90$.  Failure
occurs via the $K_1$-mode in region II and via the elastic instability in
region I and III. ({\bf B}) The nominal stress in the $x$ and $y$ directions for all points
along the $K$-cell curve in panel A. ({\bf C}) The same as panel A, except uniaxial strains are applied
in the $x'$ and $y'$ directions, which correspond to a $15$ degree rotation of the coordinate system.

\begin{figure}[htb]
\begin{center}
\includegraphics[width=\linewidth,clip= ]{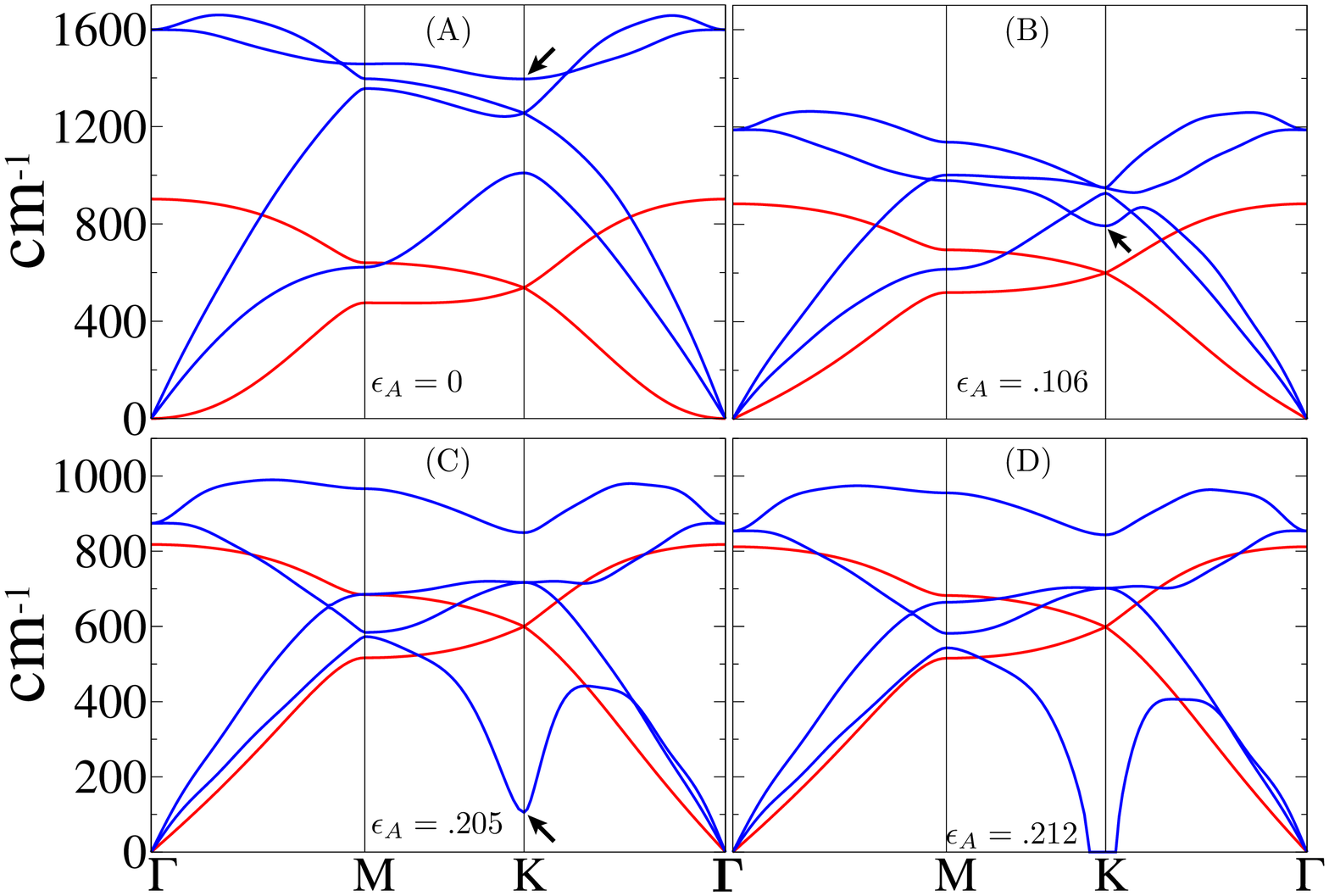}
\end{center}
\caption{}
\label{fig}
\end{figure}

\begin{figure}[htb]
\begin{center}
\includegraphics[width=\linewidth,clip= ]{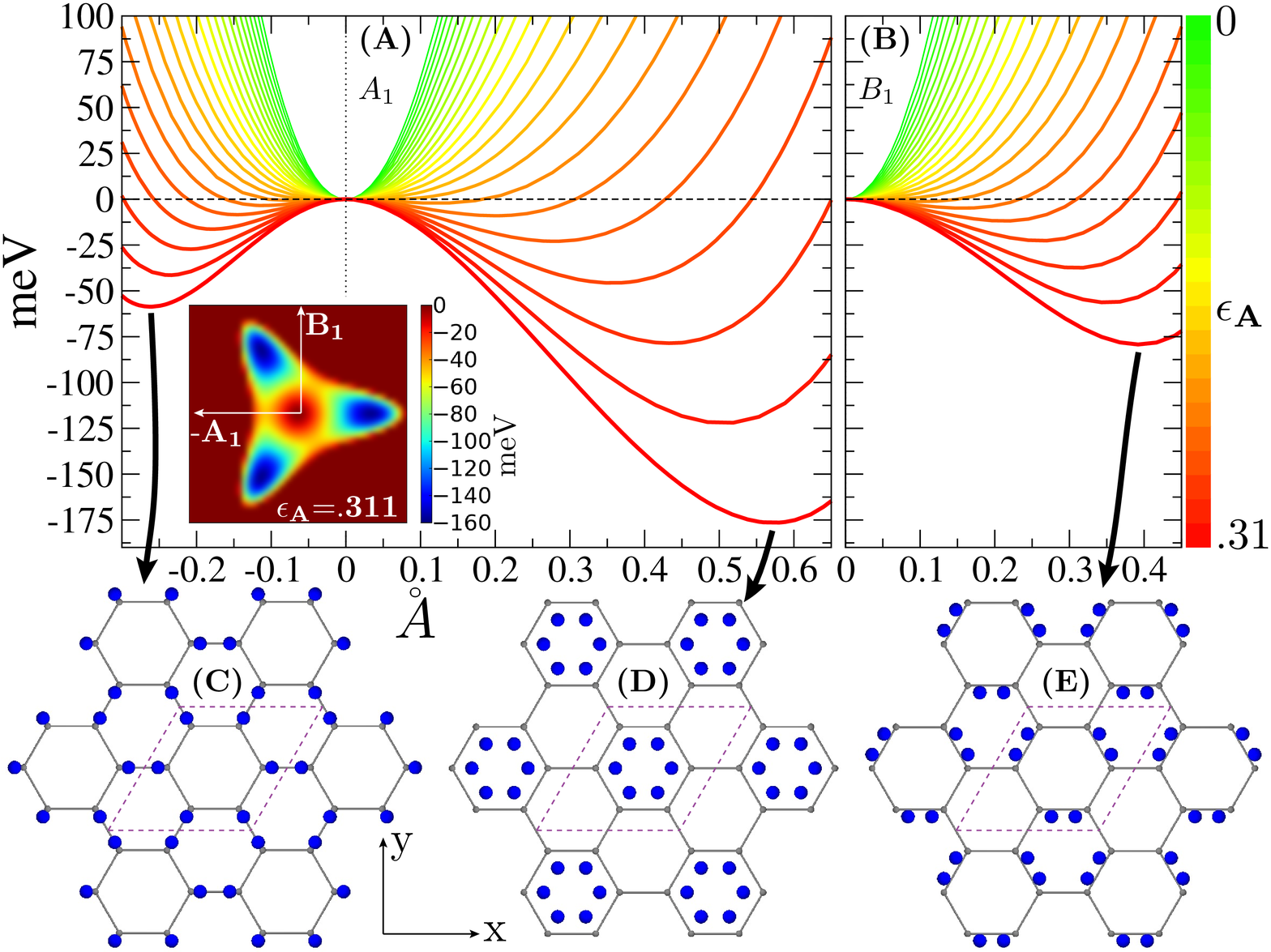}
\end{center}
\caption{}
\label{fig}
\end{figure}

\begin{figure}[htb]
\begin{center}
\includegraphics[width=\linewidth,clip= ]{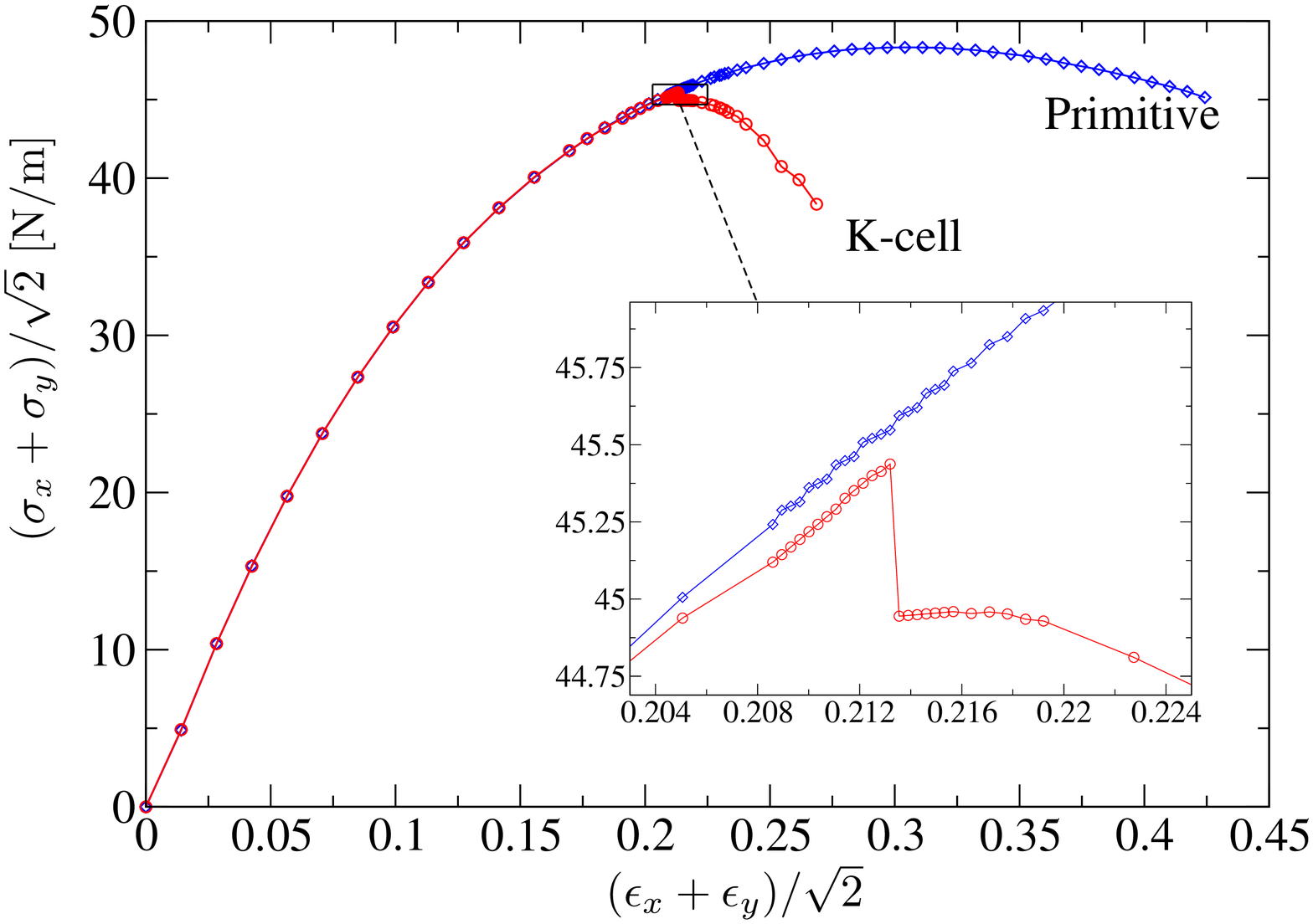}
\end{center}
\caption{}
\label{fig}
\end{figure}

\begin{figure}[htb]
\begin{center}
\includegraphics[width=\linewidth,clip= ]{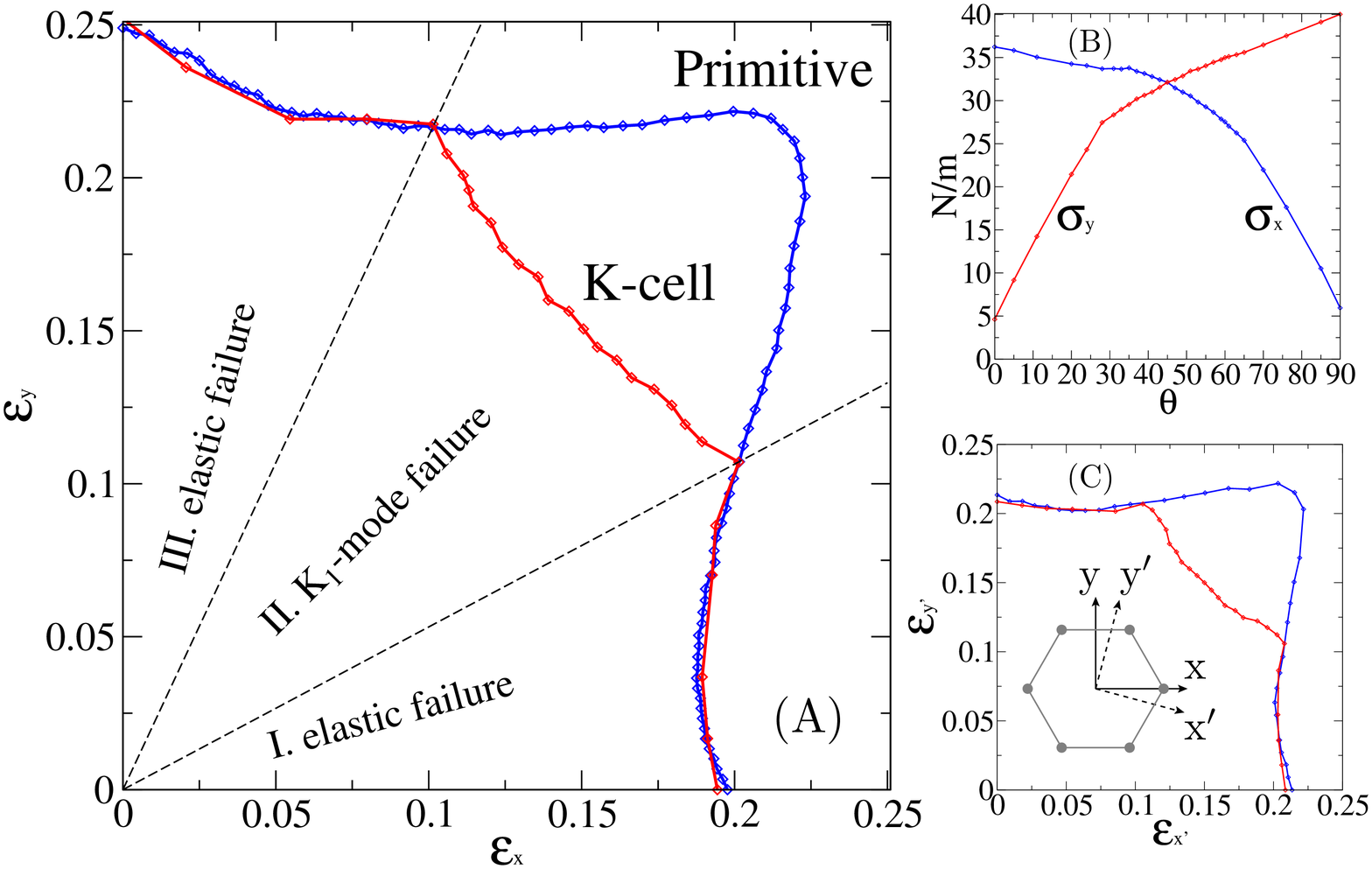}
\end{center}
\caption{}
\label{fig}
\end{figure}

\end{document}